\documentclass[11pt]{article}
\usepackage{epsfig}

\usepackage{amssymb}
\usepackage{epigraph}

\usepackage{natbib}
\usepackage{amsmath}
\usepackage{amsfonts}
\usepackage{amssymb}
\usepackage{newtxmath}   
\usepackage{float}
\usepackage{mathrsfs}
\usepackage[framemethod=TikZ]{mdframed}
\usepackage{xcolor}
\usepackage{hyperref}
\usepackage{multimedia}
\usepackage{graphicx}%
\usepackage{comment}
\usepackage{bbm}

\providecommand{\U}[1]{\protect\rule{.1in}{.1in}}

\newcommand{\beq}{\begin{equation}}
\newcommand{\eeq}{\end{equation}}
\newcommand{\bea}{\begin{eqnarray}}
\newcommand{\eea}{\end{eqnarray}}

\newtheorem{theorem}{Theorem}[section]

\newtheorem{example}[theorem]{Example}

\newcommand{\be}{\begin{equation}}
\newcommand{\ee}{\end{equation}}

\newcommand{\qed}{\nobreak \ifvmode \relax \else
      \ifdim\lastskip<1.5em \hskip-\lastskip
      \hskip1.5em plus0em minus0.5em \fi \nobreak
      \vrule height0.75em width0.5em depth0.25em\fi}

\newcounter{example}
\renewcommand{\theexample}{\thesection.\arabic{example}}

\mdfdefinestyle{bluebox}{%
    backgroundcolor=blue!8,
    linecolor=blue,
    outerlinewidth=2pt,
    leftline=false,rightline=false,
    skipabove=\baselineskip,
    skipbelow=\baselineskip,
    frametitle=\mbox{},
}
\newmdenv[%
    style=bluebox,
    settings={\global\refstepcounter{bluebox}},
    frametitlefont={\bfseries Example~\theexample\quad},
]{bluebox}
\newmdenv[%
    style=bluebox,
    frametitlefont={\bfseries Example~\quad},
]{bluebox*}

\newmdenv[%
    backgroundcolor=red!8,
    linecolor=red,
    outerlinewidth=1pt,
    roundcorner=5mm,
    skipabove=\baselineskip,
    skipbelow=\baselineskip,
]{redbox}

\usepackage[english]{babel}
\usepackage{blindtext}

\title{G-Learner and GIRL: \\
\vspace{0.3cm}
Goal Based Wealth Management with Reinforcement Learning}

\author{Matthew F. Dixon\thanks{Matthew Dixon is an Assistant Professor in the Department of Applied Math, Illinois Institute of Technology. E-mail: matthew.dixon@iit.edu.}\\
    Department of Applied Math\\
    Illinois Institute of Technology\\
    \\
    Igor Halperin\thanks{Igor Halperin is a Research Professor in Financial Engineering at NYU, and an AI Research associate at Fidelity Investments. E-mail: ighalp@gmail.com. The views presented in this paper are of the author, and do not necessarily represent the views of his employer. The standard disclaimer applies. The author thanks Lisa Huang for helpful discussions.}\\
    Fidelity Investments \&\\
    NYU Tandon School of  Engineering \\
    }
\date{February 2020}

\begin{document}
 \maketitle

\begin{abstract}
We present a reinforcement learning approach to goal based wealth management problems such as optimization of retirement plans or target dated funds. In such problems, 
an investor seeks to achieve a financial goal by making periodic investments in the portfolio while being employed, and periodically draws from the account when in retirement,
%In addition to adding or withdrawing capital %to the portfolio, the investor can also 
in addition to the ability to
re-balance the portfolio by selling and buying different assets (e.g. stocks). Instead of relying on 
 a utility of consumption, we present G-Learner: a reinforcement learning algorithm that operates %works% 
 with explicitly defined one-step rewards, does not assume a data generation process, and is suitable for noisy data. 
 Our approach is based on G-learning \citep{G-Learning} --- a probabilistic extension of the Q-learning method of reinforcement learning. 
 %By defining actions as dollar-valued %changes of asset positions, 
 In this paper, we demonstrate how G-learning, when applied to a quadratic reward and Gaussian reference policy, gives an entropy-regulated Linear Quadratic Regulator (LQR). This critical insight provides a novel and computationally tractable tool for wealth management tasks which scales to high dimensional portfolios.
%and results in 
%a Gaussian optimal policy whose mean is a %linear function of the state. 
In addition to the solution of the direct problem of G-learning, we also present a new algorithm, GIRL, that extends our goal-based G-learning approach to the setting of Inverse Reinforcement Learning (IRL) where rewards collected by the agent are not observed, and should instead be inferred.
We demonstrate that GIRL can successfully learn the reward parameters of a G-Learner agent and thus imitate its behavior. Finally, we discuss potential applications of the G-Learner and GIRL algorithms for wealth management and robo-advising.
    
\end{abstract}
. 
%We  propose a reward function whose maximization gives rise to the Conditional Value at Risk (CVaR) of a wealth growth.
% as a coherent risk metric for optimal financial planning under uncertainty.
%A one-period forward-looking wealth planning amounts to computing the CVaR with an optimal upfront investment determined by the Value at Risk (VaR) of a wealth at the end of a planning period. For a multi-period case, this produces a dynamic problem of a risk-sensitive control, which can be solved using methods of reinforcement learning.
 
%Numerically, it amounts to a sample-based (distribution-free) multi-period minimization of a risk measure given portfolio constraints. 
%Choosing CVaR (Conditional Value at Risk) as a risk metric, this can be reduced to multi-step convex optimization, enabling using of standard of-the-shelf optimization software.

%\epigraph{Keep it real.} 
%{Ali G}

\section{Introduction}

Mean-variance Markowitz optimization (MVO) \citep{Markowitz}
remains one of the most commonly used tools in wealth management. Portfolio objectives in this approach are defined in terms of expected returns and covariances of assets in the portfolio, which may not be the most natural formulation for retail investors.
Indeed, the latter typically seek specific financial goals for their portfolios. For example, a contributor to a retirement plan may demand that the value of their portfolio at the age of his or her retirement be at least equal to, or preferably larger than, some target value 
$ P_T$. 

Goal-based wealth management offers some valuable perspectives into optimal structuring of wealth management plans such as retirement plans or target date funds. The motivation for operating in terms of wealth goals can be more intuitive (while still tractable) than the classical formulation in 
terms of expected excess returns and variances. To see this, let $ V_T $ be the final 
wealth in the portfolio, and $ P_T $ be a certain target wealth level at the horizon $ T$. The goal-based wealth management approach of
\cite{Browne_1996} and \cite{Das_2018}
 uses the probability $ {\bf P} \left[ V_T  - P_T \geq 0 \right] $ of final wealth $ V_T $ to be above the target level $ P_T $ as an objective for maximization by an active portfolio management. This probability is the same as the price of a binary option on the terminal wealth $ V_T $ with strike $ P_T $: $ {\bf P} \left[ V_T - P_T \geq 0 \right]  = \mathbb{E}_t \left[ \mathbbm{1}_{V_T > P_T} \right] $. 
 Instead of a utility of wealth such as e.g. a power or logarithmic utility, this approach uses the price of this binary option as the objective function. 
This idea can also be modified by using a call option-like expectation $
\mathbb{E}_t \left[ \left( V_T - P_T \right)_{+} \right] $, instead of a binary  
option. Such an expectation quantifies how much the terminal wealth is expected to exceed the target, rather than simply providing the probability of such event\footnote{
 The problem of optimal consumption with an investment portfolio is frequently referred to as the \emph{Merton consumption problem}\index{Merton consumption problem}, after the celebrated work of Robert Merton who formulated this problem as a continuous-time optimal control problem with log-normal dynamics for asset prices \citep{Merton_1971}. As optimization in problems involving cash injections instead of cash withdrawals formally corresponds to a sign change of one-step consumption in the Merton formulation, we can collectively 
 %refer to all such problems of portfolio %optimization without self-financing %condition as Merton consumption problems.
 %We can therefore 
 refer to all types of wealth management problems involving injections or withdrawals of funds at intermediate time steps as a generalized Merton consumption problem. 
 }.

This treatment of the goal-based utility function can be implemented in a reinforcement learning (RL) framework for discrete-time planning problems. In contrast to the Merton consumption approach, RL does not require specific functional forms of the utility nor does it require that the dynamics of the assets be treated as log-normal.
Thus in theory, RL can be viewed as a data-driven extension of dynamic programming \citep{SB}. 
In practice, 
a substantial challenge with the RL framework is the curse of dimensionality --- portfolio allocation as a continuous action space Markov Decision Process (MDP) requires techniques such as deep Q-learning or other function approximation methods combined e.g. with the Least Squares Policy Iteration (LSPI) method \citep{LSPI}. The latter has exponential complexity with increasing stocks in the portfolio, and the former is cumbersome, highly data intensive, and heavily relies on heuristics for operational efficiency.
For more details, see e.g. \citep{DHB}.

In this paper, we %use%
present G-learning \citep{G-Learning} --- a probabilistic extension of Q-learning which scales to high dimensional portfolios while providing a flexible choice of utility functions. To demonstrate the utility of G-learning, we consider a general class of wealth management problems:
 optimization of a defined contribution retirement plan, where cash is injected (rather than withdrawn) at each time step. 
 In contrast to methods based on a utility of consumption, 
 we adopt a more ``RL-native'' approach by directly specifying one-step rewards. Such an approach is sufficiently general to capture other possible settings, such as e.g. a retirement plan in a decumulation (post-retirement) phase, or target based wealth management. Previously, G-learning was applied to dynamic portfolio optimization in \citep{IHIF}, while here we extend this approach to portfolio management involving cashflows at intermediate time steps.
 
 A key step in our formulation is that we define actions as absolute (dollar-valued) changes of asset positions, instead of defining them in fractional terms, as in the Merton approach \citep{Merton_1971}. This enables a simple transformation of the optimization problem into an unconstrained optimization problem, and provides a semi-analytical solution for a particular choice of the reward function. 
 As will be shown below, this approach offers a tractable setting for both the direct reinforcement learning problem of learning the optimal policy which maximizes the total reward, and its \emph{inverse} problem where we observe actions of a financial agent but not the rewards received by the agent. 
 Inference of the reward function from observations of states and actions of the agent is the objective of Inverse Reinforcement Learning (IRL). After we present \emph{G-Learner} --- a G-learning algorithm for the direct RL problem, we will introduce \emph{GIRL} (G-learning IRL) --- a framework for inference of  rewards of financial agents
 that are ``implied'' by their observed behavior. The two practical algorithms, G-Learner and GIRL, can be used either separately or in a combination, and we will discuss their potential joint applications for wealth management and robo-advising. 
 
 The paper is organized as follows.
 In Section \ref{sect_G_learning}, we introduce G-learning and explain how it generalizes the more well known Q-learning method for reinforcement learning.
 Section \ref{sect:dcp} introduces the problem of portfolio optimization for a defined contribution retirement plan. Then in Section \ref{sect:G_learning_wealth}, we present the G-Learner: a G-learning algorithm for portfolio optimization with cash injection and consumption. The GIRL algorithm for performing IRL of financial agents is introduced in Section \ref{sect_GIRL}.
 Section \ref{sect:num} presents the results of our implementation and demonstrates the ability of G-learner to scale to high dimensional portfolio optimization problems, and the ability of GIRL to make inference of the reward function of a G-Learner agent.
 Section \ref{sect:concl} concludes with ideas for future developments in G-learning for wealth management and robo-advising.

\section{G-learning}
\label{sect_G_learning}

In this section, we provide a short but self-contained overview of G-learning as a probabilistic extension of the popular Q-learning method in reinforcement learning.
We assume some familiarity with constructs in dynamic programming and reinforcement learning, see e.g. \citep{SB}, or \citep{DHB} for a more finance-focused introduction. In particular, we assume that the reader is familiar with the notions of value function, action-value function, and the Bellman optimality equations. Familiarity with Q-learning is desirable but not critical for understanding this section, however for the benefit of the informed reader, a short informal summary of the differences is as follows:
\begin{itemize}
\item \emph{Q-learning} is an off-policy RL method with a \emph{deterministic} policy.

\item \emph{G-Learning} is an off-policy RL method with a \emph{stochastic} policy.  G-learning can be considered as an entropy-regularized Q-learning, which may be suitable when working with noisy data. Because  G-learning operates with stochastic policies, it amounts to a generative RL model.
\end{itemize}

\subsection{Bellman optimality equation}

More formally, let $ {\bf x}_t $ be a state vector for an agent that summarizes the knowledge of the environment that the agent needs in order to perform an action $ {\bf a}_t $ at time step $ t $\footnote{Here we assume a discrete-time setting where time $ t $ is measured in terms of integer-valued number of elementary time steps $ \Delta t$.}. 
Let $ \hat{R}_{t} ({\bf x}_{t}, {\bf a}_{t} )$ be a random reward collected by the agent for taking action $ {\bf a}_{t} $ at time $ t $ when the state of the environment %world%
is $ {\bf x}_{t}$.
Assume that all future actions $ {\bf a}_{t}$
for future time steps are determined according to a policy $ \pi({\bf a}_{t} | {\bf x}_{t}) $ which specifies which action 
$ {\bf a}_{t}$ to take when the environment is in state  $ {\bf x}_t $. 
We note that policy $ \pi $ can be deterministic as in Q-learning, or stochastic as in G-learning, as we will discuss below.

For a given policy $ \pi $, the expected value of cumulative reward with a discount factor $ \gamma$, conditioned on the current state ${\bf x}_{t} $, defines the value function
\beq
\label{V_star}
V_t^{\pi} ( {\bf x}_t) :=  \mathbb{E}_t^{\pi} \left[ \left. \sum_{t'=t}^{T-1} \gamma^{t'-t} \hat{R}_{t'} ({\bf x}_{t'}, {\bf a}_{t'} ) \right| {\bf x}_t \right].
\eeq
Here $ \mathbb{E}_t^{\pi}$ stands for the expectation of future states and actions, conditioned on the current state $ {\bf x}_t $
and policy $ \pi$. 

Let $ \pi^{\star}$ be the {\it optimal} policy, i.e. the policy that maximizes the total reward. This policy corresponds to the {\it optimal} value function, denoted 
$ V_t^{\star} ( {\bf x}_t) $. The latter satisfies 
the Bellman optimality equation 
(see e.g. \citep{SB})
\beq
\label{Bellman_V}
V_t^{\star} ( {\bf x}_t) = \max_{ {\bf a}_t } \, \hat{R}_t ({\bf x}_t, {\bf a}_t ) + \gamma \mathbb{E}_{ 
t, {\bf a}_t} \left[  V_{t+1}^{\star} ( {\bf x}_{t+1}) \right]. % := \left( \mathcal{T} V_t^{\star} \right) ({\bf x}_t)
\eeq 
%where $ \mathcal{T} $ is the Bellman operator. 
Here $ \mathbb{E}_{ 
t, {\bf a}_t} \left[  \cdot \right] $ stands for an expectation conditional on the current state $ {\bf x}_t$ and action $ {\bf a}_t$.
The optimal policy $ \pi^{\star} $ can be obtained from $ V^{\star} $ as follows:
\beq
\label{pi_star_V_star}
 \pi_t^{\star} ({\bf a}_t | {\bf x}_t ) = \arg \max_{ {\bf a}_t} \,  \hat{R}_t ({\bf x}_t, {\bf a}_t) +  \gamma \mathbb{E}_{ 
t, {\bf a}_t } \left[  V_{t+1}^{\star} ( {\bf x}_{t+1}) \right]. 
\eeq
The goal of Reinforcement Learning (RL) is to solve the Bellman optimality equation based on samples of data. Assuming that an optimal value function is found by means of RL, solving for the optimal policy $ \pi^{\star} $ takes another optimization problem as formulated in Eq.(\ref{pi_star_V_star}).  

\subsection{Entropy-regularized Bellman optimality equation} \label{sect:ent-bellman}

%Following \cite{Dai}, 
Let us begin by reformulating the Bellman optimality equation using a Fenchel-type representation:
\beq
\label{V_star_Fenchel}
 V_t^{\star} ( {\bf x}_t)  = \max_{ \pi(\cdot| y) \in \mathcal{P}} \sum_{{\bf a}_t \in \mathcal{A}_t }  \pi ({\bf a}_t | {\bf x}_t )
\left(   \hat{R}_t ({\bf x}_t, {\bf a}_t) +  \gamma \mathbb{E}_{ 
t, {\bf a}_t } \left[  V_{t+1}^{\star} ( {\bf x}_{t+1}) \right] \right).
\eeq
Here $ \mathcal{P} = \left\{ \pi: \, \pi \geq 0, \mathbbm{1}^T \pi = 1 \right\} $ denotes a set of all valid distributions. Eq.(\ref{V_star_Fenchel}) is equivalent to the original Bellman optimality equation (\ref{Bellman_V}), because for any $  x \in \mathbb{R}^n $, we have $ \max_{i \in \{ 1, \ldots, n \} } x_i = 
\max_{\pi \geq 0, || \pi || \leq 1 }  \pi^T x $.  Note that while we use discrete notations for simplicity of presentation, all formulae below can be equivalently expressed in continuous notations by replacing sums by integrals. For brevity, we will denote the expectation $ \mathbb{E}_{ 
{\bf x}_{t+1}| {\bf x}_t, {\bf a}_t } \left[ \cdot \right] $ as $ \mathbb{E}_{t, {\bf a}} \left[ \cdot \right] $ in what follows.

The one-step {\it information cost} of a learned policy $ \pi ( {\bf a}_t | {\bf x}_t) $ relative to a reference policy $ \pi_0( {\bf a}_t | {\bf x}_t) $  is defined as follows \citep{G-Learning}:
\beq
\label{info_cost_ch10}
g^{\pi} ({\bf x}_t, {\bf a}_t ) := \log \frac{  \pi ( {\bf a}_t | {\bf x}_t) }{ \pi_0 ( {\bf a}_t | {\bf x}_t) }.
\eeq
Its expectation with respect to the policy $ \pi $ is the Kullback-Leibler (KL) divergence of $ \pi(\cdot|  {\bf x}_t) $ and $ \pi_0( \cdot |  {\bf x}_t) $:
\beq
\label{Eg_KL}
\mathbb{E}_{\pi} \left[ \left. g^{\pi} ({\bf x}, {\bf a} ) \right| {\bf x}_t \right] = KL[ \pi || \pi_0] ({\bf x}_t) := 
%\int d {\bf a}_t
\sum_{{\bf a}_t } 
\pi ( {\bf a}_t | {\bf x}_t)  
\log \frac{ \pi ( {\bf a}_t | {\bf x}_t) }{\pi_0 ( {\bf a}_t | {\bf x}_t) }. 
\eeq
The total discounted information cost for a trajectory is defined as follows:
\beq
\label{I_pi}
I^{\pi}({\bf x}_t ) := \sum_{t'=t}^{T} \gamma^{t'-t} \mathbb{E}_t^{\pi} \left[ \left. g^{\pi} ({\bf x}_{t'}, {\bf a}_{t'} ) \right| {\bf x} _t \right].
\eeq
The {\it free energy} function $ F_t^{\pi} ({\bf x}_t) $ is defined as the value function (\ref{V_star_Fenchel}) augmented by the information cost penalty (\ref{I_pi}) which is added using a regularization parameter $ 1/\beta$:
%  and the cost of the self-financing constraint (\ref{self_fin_dc}):
\beq
\label{F_pi_ch10}
F_t^{\pi}({\bf x}_t) 
  :=   V_t^{\pi} ( {\bf x}_t) - \frac{1}{\beta}  I^{\pi} ( {\bf x}_t)  %\nonumber \\
%& =&
= \sum_{t'=t}^{T} \gamma^{t'-t} 
\mathbb{E}_t^{\pi} \left[  \hat{R}_{t'} ({\bf x}_{t'}, {\bf a}_{t'} ) - \frac{1}{\beta}  g^{\pi} ({\bf x}_{t'}, {\bf a}_{t'} ) 
\right].  
\eeq
The free energy, $ F_t^{\pi}({\bf x}_t)  $, 
is the entropy-regularized value function, where the amount of regularization can be tuned to the level of noise in the data. 
The regularization parameter $ \beta $  in Eq.(\ref{F_pi_ch10}) 
controls a trade-off between reward optimization and proximity of the optimal policy to the reference policy, and is 
often referred to as the ``inverse temperature'' parameter, using the analogy between Eq.(\ref{F_pi_ch10}) and free energy in physics, see e.g. \citep{DHB}. 
The reference
policy, $ \pi_0 $, provides a ``guiding hand'' in the stochastic policy optimization process that we now describe. 

A Bellman equation for the free energy function $ F_t^{\pi} ({\bf x}_t) $ is obtained from Eq.(\ref{F_pi_ch10}):
\beq
\label{Bellman_F}
F_t^{\pi}({\bf x}_t)  =    \mathbb{E}_{ {\bf a}|y}  \left[ \hat{R}_{t} ({\bf x}_{t}, {\bf a}_{t} )
 - \frac{1}{\beta}  g^{\pi} ({\bf x}_{t}, {\bf a}_{t})  + 
\gamma \mathbb{E}_{t, {\bf a}}  \left[ F_{t+1}^{\pi}({\bf x}_{t+1})  \right]  \right].
\eeq
For a finite-horizon setting with a terminal reward $ \hat{R}_{T} ({\bf x}_{t}, {\bf a}_{T} )$, Eq.(\ref{Bellman_F}) should be supplemented by a terminal condition 
\beq
\label{F_pi_ch10_T}
F_T^{\pi}({\bf x}_t) =  \hat{R}_{T} ({\bf x}_{t}, {\bf a}_{T}^{\star} ) 
\eeq
where the final action $ {\bf a}_{T}^{\star} $ maximizes the terminal reward $ \hat{R}_{T}$ for the given terminal state $ {\bf x}_{T} $.
Eq.(\ref{Bellman_F}) can be viewed as a soft probabilistic relaxation of the Bellman equation for the value function, with the KL information cost penalty 
(\ref{info_cost_ch10}) as a regularization controlled by the inverse temperature $ \beta $. In addition to such a regularized value function (free energy), we will next introduce an entropy regularized Q-function.

\subsection{G-function: an entropy-regularized Q-function}

Similar to the action-value function, we define the state-action free energy function $ G^{\pi} ( {\bf x}, {\bf a}) $ as \citep{G-Learning}
\bea
\label{G_fun_ch10}
G_t^{\pi} ( {\bf x}_t, {\bf a}_t) 
&=& \hat{R}_t ({\bf x}_{t}, {\bf a}_{t} ) 
 +  \gamma  \mathbb{E} \left[  \left. F_{t+1}^{\pi} ( {\bf x}_{t+1}) \right|  {\bf x}_t, {\bf a}_t \right]   \\  
&=&  \hat{R}_t ({\bf x}_{t}, {\bf a}_{t} ) 
   + \gamma  \mathbb{E}_{t, {\bf a}} \left[  \sum_{t'=t+1}^{T} \gamma^{t'-t-1} \left(   \hat{R}_{t'} ({\bf x}_{t'}, {\bf a}_{t'} ) - 
   \frac{1}{\beta} g^{\pi} ({\bf x}_{t'}, {\bf a}_{t'} ) \right)  \right] \nonumber \\
&=&  \mathbb{E}_{t, {\bf a}_t} \left[   \sum_{t'=t}^{T}  \gamma^{t'-t} \left(   \hat{R}_{t'} ({\bf x}_{t'}, {\bf a}_{t'} ) - 
   \frac{1}{\beta} g^{\pi} ({\bf x}_{t'}, {\bf a}_{t'} ) \right)  \right] \nonumber, 
\eea
where in the last equation we used the fact that the first action $ {\bf a}_t $ 
in the G-function is fixed, and hence $ g^{\pi} ({\bf x}_{t}, {\bf a}_{t} ) = 0 $ when we condition on $ {\bf a}_t$.

If we now compare this expression with Eq.(\ref{F_pi_ch10}), we obtain the relation between the G-function and the free energy
$ F_t^{\pi}({\bf x}_t ) $:
\beq
\label{G_F_ch10}
F_t^{\pi}({\bf x}_t ) = \sum_{ {\bf a}_t} \pi ( {\bf a}_t | {\bf x}_t) \left[ G_t^{\pi} ( {\bf x}_t, {\bf a}_t) -  \frac{1}{\beta}
\log \frac{  \pi ( {\bf a}_t | {\bf x}_t) }{ \pi_0 ( {\bf a}_t | {\bf x}_t) } \right].
\eeq
%To maximize the free energy under the self-financing constraint of Eq.(\ref{opt_2}), we add it to (\ref{G_F}) using the method of Lagrange multipliers to form an augmented free energy $ \hat{F}_t^{\pi} ({\bf w}_t, \xi ) $:
%\beq
%\label{G_F_xi}
%\hat{F}_t^{\pi}({\bf w}_t, \xi ) = \sum_{ {\bf a}_t} \pi ( {\bf a}_t | {\bf w}_t) \left[ G_t^{\pi} ( {\bf w}_t, {\bf a}_t) -  \frac{1}{\beta}
%\log \frac{  \pi ( {\bf a}_t | {\bf w}_t) }{ \pi_0 ( {\bf a}_t | {\bf w}_t) } 
%- \xi \left( {\bf b}^T {\bf a}_t +  \tilde{\Psi_t}^T   {\bf W} +  \frac{\hat{\eta}}{2} L_{max}  \right)
%\right]
%\eeq
%where $ \xi $ is a Lagrange multiplier. 
This functional is maximized by the following distribution $ \pi ( {\bf a}_t | {\bf x}_t) $:
\bea
\label{pi_from_F_ch10}
 && \pi ( {\bf a}_t | {\bf x}_t) = \frac{1}{Z_t} \pi_0 ( {\bf a}_t | {\bf x}_t) e^{ \beta G_t^{\pi} ( {\bf x}_t, {\bf a}_t) }  \\
 && Z_t = \sum_{{\bf a}_t} \pi_0 ( {\bf a}_t | {\bf x}_t) e^{ \beta G_t^{\pi} ( {\bf x}_t, {\bf a}_t) } \nonumber. 
 \eea 
The free energy (\ref{G_F_ch10}) evaluated at the optimal solution (\ref{pi_from_F_ch10}) becomes
\beq
\label{F_opt_ch10}
F_t^{\pi}({\bf x}_t ) =  \frac{1}{\beta} \log Z_t =  \frac{1}{\beta} \log \sum_{{\bf a}_t} \pi_0 
( {\bf a}_t | {\bf x}_t) e^{ \beta G_t^{\pi} ( {\bf x}_t, {\bf a}_t)  }. 
\eeq
Using Eq.(\ref{F_opt_ch10}), the optimal action policy can be written as follows :
\beq
\label{pi_opt_F_ch10}
 \pi ( {\bf a}_t | {\bf x}_t) = \pi_0 ( {\bf a}_t | {\bf x}_t) e^{ \beta \left(G_t^{\pi} ( {\bf x}_t, {\bf a}_t) - F_t^{\pi}({\bf x}_t ) 
  \right) }. 
 \eeq 
 Eqs.(\ref{F_opt_ch10}), (\ref{pi_opt_F_ch10}), along with the first form of Eq.(\ref{G_fun_ch10}) repeated here for convenience:
 \beq
 \label{G_from_F_ch10_2}
 G_t^{\pi} ( {\bf x}_t, {\bf a}_t) 
 = \hat{R}_{t}({\bf x}_{t}, {\bf a}_{t} ) 
 +  \gamma  \mathbb{E}_{t, {\bf a}} \left[  \left. F_{t+1}^{\pi} ( {\bf x}_{t+1}) \right|  {\bf x}_t, {\bf a}_t \right], 
 \eeq
 constitute a system of equations 
 for G-learning \citep{G-Learning} that should be solved self-consistently for $  \pi ( {\bf a}_t | {\bf x}_t) $, $ G_t^{\pi} ( {\bf x}_t, {\bf a}_t) $ and $ F_t^{\pi}({\bf x}_t ) $ by backward recursion for $ t = T-1, \ldots, 0 $,
 with terminal conditions 
\bea
\label{terminal_G_F}
&& G_T^{\pi} ( {\bf x}_t, {\bf a}_T^{\star}) 
 = \hat{R}_{T}({\bf x}_{t}, {\bf a}_{T}^{\star} ) \\
 &&  F_T^{\pi}({\bf x}_t) = G_T^{\pi} ( {\bf x}_t, {\bf a}_T^{\star}) 
 =  \hat{R}_{T}({\bf x}_{t}, {\bf a}_{T}^{\star} ). \nonumber 
 \eea 
 %Eqs.(\ref{F_opt_ch10}, \ref{pi_opt_F_ch10}, %\ref{G_from_F_ch10_2})
 We will next show how G-learning can be implemented in the context of (direct) reinforcement learning.
 
\subsection{G-learning}
\label{sect:G-learning} 
 
 In the RL setting when rewards are observed, the system Eqs.(\ref{F_opt_ch10}, \ref{pi_opt_F_ch10}, \ref{G_from_F_ch10_2}) can be reduced to one non-linear equation.
 Substituting the augmented free energy (\ref{F_opt_ch10}) into Eq.(\ref{G_from_F_ch10_2}), we obtain
 \beq
 \label{soft_Q_ch10}
 G_t^{\pi} ( {\bf x}, {\bf a}) 
  =  \hat{R} ({\bf x}_{t}, {\bf a}_{t} ) +  \mathbb{E}_{t, {\bf a}} \left[  
   \frac{\gamma}{\beta}  \log \sum_{{\bf a}_{t+1}} \pi_0 
( {\bf a}_{t+1} | {\bf x}_{t+1}) e^{ \beta G_{t+1}^{\pi} ( {\bf x}_{t+1}, {\bf a}_{t+1})  } \right]. 
\eeq
This equation provides a soft relaxation of the Bellman optimality equation for the action-value Q-function, with the G-function defined in
Eq.(\ref{G_fun_ch10}) being an entropy-regularized Q-function \citep{G-Learning}. 
 The "inverse-temperature" parameter $ \beta $ in Eq.(\ref{soft_Q_ch10}) 
 determines the strength of entropy regularization. In particular, if we 
 take a ``zero-temperature" limit $ \beta \rightarrow \infty $, we recover the original Bellman optimality equation for the Q-function.  Because the last term in (\ref{soft_Q_ch10}) approximates the  $ \max(\cdot) $ function when $ \beta $ is large but finite, for a particular choice of a uniform reference distribution $ \pi_0 $, Eq.(\ref{soft_Q_ch10}) is known in the literature as ``soft Q-learning''.
 
 For finite values $ \beta < \infty $, in a setting of Reinforcement Learning with observed rewards, Eq.(\ref{soft_Q_ch10}) can be used to specify {\it G-learning}
  \citep{G-Learning}: an off-policy time-difference (TD) algorithm that generalizes Q-learning to noisy environments where an entropy-based regularization is appropriate\index{G-learning}.%might be needed%. 
  
  The G-learning algorithm of \cite{G-Learning} was specified in a tabulated setting where both the state and action space are finite.
 In our case, we model MDPs in
 %deal with% 
 high-dimensional continuous state and action spaces. 
 Respectively, we cannot rely on a tabulated G-learning, and need to specify a functional form of the action-value function, or use a non-parametric function approximation such as a neural network to represent its values. An additional challenge is to compute a multidimensional integral (or a sum) over all next-step actions in Eq.(\ref{soft_Q_ch10}). Unless a tractable paramet{\color{blue}e}rization is used for $ \pi_0 $ and 
 $ G_t $, repeated numerical integration of this integral
 can substantially slow down the learning.

 To summarize, G-learning is an off-policy, generative reinforcement learning algorithm with a stochastic policy. In contrast to Q-learning, which produces deterministic policies, G-learning
 generally produces stochastic policies, while the deterministic Q-learning policies are recovered in a zero-temperature limit $ \beta \rightarrow \infty $. In the next section, we will build an approach to goal-based wealth management based on G-learning. Later in this paper, we will also consider applications of G-learning for Inverse Reinforcement Learning (IRL). 
 
\section{Portfolio optimization for a defined contribution retirement plan} \label{sect:dcp}

Let us begin by considering a simplified model for retirement planning. We assume a discrete-time process with $ T $ steps, so that $ T $ is the (integer-valued) time horizon. The investor/planner keeps the wealth in $ N $ assets, with $ {\bf x}_t$ being the vector of dollar values of positions in different assets at time $ t $, and $ {\bf u}_t $ being the vector of changes in these positions. 
We assume that the first asset with $ n = 1 $ is a risk-free bond, and other assets are risky, with uncertain returns $ {\bf r}_t$ whose expected values are $ \bar{\bf r}_t$. The covariance matrix of return is $ {\bf \Sigma}_r $ of size $ (N-1)\times (N-1)$. 
%Note that our notation in this section is %different from the previous section where $ %{\bf x}_t$ was used to denote a vector of {\it %risky} asset holding values.

Optimization of a 
retirement plan involves optimization of both regular contributions to the plan and asset allocations. Let $ c_t $ be a cash installment in the plan at time $ t $.
The pair $ ( c_t, {\bf u}_t )$ can thus be considered the action variables in a dynamic optimization problem corresponding to the retirement plan.  

We assume that at each time step $ t $, there is a pre-specified target value $ \hat{P}_{t+1} $ of a portfolio at time $ t + 1$. 
We assume that the target value  $ \hat{P}_{t+1} $ at step $ t $ 
exceeds the next-step value $ V_{t+1} = (1 + {\bf r}_t) ({\bf x}_t + 
{\bf u}_t)$ of the portfolio,
and we seek to impose a penalty for under-performance relative to this target.
%We want to maximize the expected difference between %the value of portfolio 
%$ V_{t+1} = (1 + {\bf r}_t) ({\bf x}_t + 
%{\bf u}_t)$ and the target value $ \hat{P}_{t+1} $ %provided that $  V_{t+1} > 
%\hat{P}_{t+1}$. 
To this end, we can consider the following expected reward for time step $ t $:
\beq
\label{one_step_R_rp}
R_t({\bf x}_t, {\bf u}_t, c_t) = - c_t - \lambda \mathbb{E}_t \left[ 
\left( \hat{P}_{t+1} -
(1+{\bf r}_t)( {\bf x}_t + {\bf u}_t ) 
\right)_{+} \right] 
- {\bf u}_t^T {\bf \Omega} {\bf u}_t. 
\eeq
Here the first term is due to an installment of amount $ c_t$ at the beginning of time period $ t $, the second term is the expected negative reward from the end of the period for under-performance relative to the target, and the third term approximates transaction costs by a convex functional with the parameter matrix $ {\bf \Omega} $, and serves as a $L_2$ regularization.  

The one-step reward (\ref{one_step_R_rp}) is inconvenient to work with due to the rectified non-linearity $ (\cdot)_{+} := \max(\cdot, 0)$ under the expectation. Another problem is that decision variables 
$ c_t $ and $ {\bf u}_t $ are not independent but rather satisfy the following constraint 
\beq
\label{constaint_rp}
\sum_{n=1}^{N} u_{tn} = c_t,
\eeq
which simply means that at every time step,
the total change in all positions should equal the cash installment $ c_t $ at this time. 

We therefore modify the one-step reward (\ref{one_step_R_rp}) in two ways: we replace the first term using Eq.(\ref{constaint_rp}), and approximate the rectified non-linearity by a quadratic function. The new one-step reward is
\beq
\label{one_step_R_rp2}
R_t({\bf x}_t, {\bf u}_t) = - \sum_{n=1}^{N} u_{tn}
- \lambda \mathbb{E}_t \left[ \left(
\hat{P}_{t+1} - 
(1+{\bf r}_t)( {\bf x}_t + {\bf u}_t ) 
\right)^2 \right] 
- {\bf u}_t^T {\bf \Omega} {\bf u}_t. 
\eeq
The new reward function (\ref{one_step_R_rp2}) is attractive on two counts. First, it explicitly resolves the constraint (\ref{constaint_rp}) between the cash injection $ c_t $ and portfolio allocation decisions, and thus converts the initial constrained optimization problem into an unconstrained one. We remind the reader that this differs from the Merton model where allocation variables are defined as fractions of the total wealth, and thus are constrained by construction. The approach based on dollar-measured actions both reduces the dimensionality of the optimization problem, and makes it unconstrained. When the unconstrained optimization problem is solved, the optimal contribution $ c_t $ at time $ t $ can be obtained from Eq.(\ref{constaint_rp}).

The second attractive feature of the reward (\ref{one_step_R_rp2}) is that it is quadratic in actions $ {\bf u}_t $, and is therefore highly tractable. On the other hand, the well known disadvantage of quadratic rewards (penalties) is that they are symmetric, and penalize both scenarios $ V_{t+1} \gg \hat{P}_{t+1} $ and 
$ V_{t+1} \ll \hat{P}_{t+1} $, while in fact we only want to penalize the second class of scenarios. To mitigate this drawback, we can consider target values $ \hat{P}_{t+1} $ that are considerably higher than the time-$t$ expectation of the next-period portfolio value. %In what follows we assume this is the case, %otherwise the value of $  \hat{P}_{t+1}$ can %be arbitrary. 
For example, one simple choice could be to set the target portfolio as a linear combination of a portfolio-independent benchmark $ B_t $ and the current portfolio growing with a fixed rate $ \eta $:
\beq
\label{target_portf}
 \hat{P}_{t+1} = (1-\rho)  B_t +  \rho \eta \, 
 \bf{1}^T {\bf x}_t, 
\eeq
where $ 0 \leq \rho \leq 1 $ is a relative weight of the portfolio-independent and 
portfolio-dependent terms, and $ \eta > 1 $ is a parameter that defines the desired growth rate of the current portfolio whose value is 
$  \bf{1}^T {\bf x}_t $. For a sufficiently large values of $ B_t $ and $ \eta$, such a target portfolio would be well above the current portfolio at all times, and thus would serve as a reasonable proxy to the asymmetric measure (\ref{one_step_R_rp}).
The advantage of such a parameterization of the target portfolio is that both the ``desired growth'' parameter $ \eta $ and the mixture parameter $ \rho $ can be learned from an observed behavior of a financial agent in the setting of Inverse Reinforcement Learning (IRL), as we will discuss in Sec.~\ref{sect_GIRL}. In what follows, we use Eq.(\ref{target_portf}) as our specification of the target portfolio.

We note that a quadratic loss specification
relative to a target time-dependent wealth level is a popular choice in the recent literature on wealth management. One example is provided by \cite{Lin_Zeng_2019} who develop a dynamic optimization approach with a similar squared loss function for a defined contribution retirement plan. 
A similar approach which relies on 
a direct specification of a reward based on a target portfolio level is known as ``goal-based wealth management'' \citep{Browne_1996, Das_2018}.

The square loss reward specification is very convenient, as it allows one to construct optimal policies semi-analytically. Here we will demonstrate how to
build a semi-analytical scheme for computing
optimal stochastic consumption-investment policies for a retirement plan --- the method is sufficiently general for either a 
cumulation or de-cumulation phase. For other specifications of rewards, numerical optimization and function approximations (e.g. neural networks) would be required.   

The expected reward (\ref{one_step_R_rp2}) can be written in a more explicit quadratic form if we denote asset returns as $ {\bf r}_t = \bar{\bf r}_t + \tilde{\bf \varepsilon}_t $ where the first component $ \bar{r}_0(t) = 
r_f $ is the risk-free rate (as the first asset is risk-free), and 
$ \tilde{\bf \varepsilon}_t = (0, {\bf \varepsilon}_t $) where 
$ {\bf \varepsilon}_t $ is an idiosyncratic noise with covariance 
$ {\bf \Sigma}_r$ of size $ (N-1) \times (N-1)$. Substituting this expression in 
Eq.(\ref{one_step_R_rp2}), we obtain
\bea
\label{one_step_R_rp3}
R_t({\bf x}_t, {\bf u}_t)  
&=&
- \lambda \hat{P}_{t+1}^2 
- {\bf u}_t^T \mathbbm{1}
+ 2 \lambda 
\hat{P}_{t+1} ({\bf x}_t + 
\bf{u}_t)^T ( 1 + \bar{\bf r}_t)
%\nonumber \\
%& -&
-
\lambda \left( {\bf x}_t + {\bf u}_t \right)^T \hat{\Sigma}_t \left( {\bf x}_t + {\bf u}_t \right)
-  {\bf u}_t^T {\bf \Omega} {\bf u}_t 
\nonumber \\
&= & {\bf x}_t^T {\bf R}_t^{(xx)}{\bf x}_t
+ {\bf u}_t^T {\bf R}_t^{(ux)}{\bf x}_t
+ {\bf u}_t^T {\bf R}_t^{(uu)}{\bf u}_t
+ {\bf x}_t^T {\bf R}_t^{(x)}
+ {\bf u}_t^T {\bf R}_t^{(u)}
+ R_t^{(0)}
%&=& - \lambda \eta^2 \rho^2 {\bf x}_t^T {\bf %1} \bf{1}^T {\bf x}_t -  
%{\bf u}_t^T \mathbbm{1}
%+ 2 \lambda 
%\eta \rho ({\bf x}_t + 
%\bf{u}_t)^T ( 1 + \bar{\bf r}_t) {\bf 1}^T 
%{\bf x}_t \nonumber \\
%& -& 
%\lambda \left( {\bf x}_t + {\bf u}_t \right)^T %\hat{\Sigma}_t \left( {\bf x}_t + {\bf u}_t %\right)
%-  {\bf u}_t^T {\bf \Omega} {\bf u}_t \\
%&-& (1-\rho)^2 \lambda B_{t}^2 - 
%2 \lambda \eta \rho (1 - \rho) B_t {\bf x}_t^T% 
%{\bf 1} 
%+ 2 \lambda (1 - \rho) B_t \left( {\bf x}_t + %{\bf u}_t \right)^T ( 1 + \bar{\bf r}_t)
\nonumber 
\eea
where 
\bea
\label{Sigma_tilde_rp}
&& \hat{\bf \Sigma}_t  = 
 \left[ \begin{array}{cc}
 0   &  {\bf 0} \\
 {\bf 0} &  {\bf \Sigma}_r 
 \end{array} \right]
+
(1+ \bar{\bf r}_t)(1 + \bar{\bf r}_t)^T
\nonumber \\
&& {\bf R}_t^{(xx)} = - \lambda \eta^2 \rho^2 {\bf 1} \bf{1}^T + 
2 \lambda \eta \rho (1 + \bar{\bf r}_t) {\bf 1}^T - \lambda \hat{\bf \Sigma}_t
\nonumber \\
&& {\bf R}_t^{(ux)} = 2 \lambda \eta \rho (1 + \bar{\bf r}_t) {\bf 1}^T 
- 2 \lambda \hat{\bf \Sigma}_t
\nonumber \\
&& {\bf R}_t^{(uu)} = -  \lambda \hat{\bf \Sigma}_t 
- {\bf \Omega} \nonumber \\
&& {\bf R}_t^{(x)} = - 2 \lambda \eta \rho (1 - \rho) B_t 
{\bf 1} + 2 \lambda (1 - \rho) B_t 
( 1 + \bar{\bf r}_t) \nonumber \\
&& {\bf R}_t^{(u)} = - {\bf 1} + 
2 \lambda  (1 - \rho) B_t 
( 1 + \bar{\bf r}_t) \\
&&  R_t^{(0)} = - (1-\rho)^2 \lambda B_{t}^2
\nonumber 
\eea
%Thus far we have considered how to deal with a %portfolio with periodic cash installments $ %c_t$. 
Assuming that the expected returns 
$ \bar{\bf r}_t$, covariance matrix $ {\bf \Sigma}_r$ and the benchmark $ B_t $ are fixed, the vector of free parameters defining the reward function is thus $ \theta := (\lambda, \eta,
\rho, \Omega)$.
 
%Because
%this approach is related to allocation decision variables by the constraint 
%(\ref{constaint_rp}), the resulting 
%quadratic reward (\ref{one_step_R_rp3}) has a similar quadratic structure as the linear LQR reward (\ref{exp_R_LQR}) for a self-financing portfolio:

\begin{comment}
The quadratic one-step reward (\ref{one_step_R_rp3}) has a similar structure to
the rewards we considered in the previous section, see e.g. Eq.(\ref{exp_R_LQR}).
In contrast to the setting in 
Sect.~\ref{sect:Zero_friction}, instead of a self-financing 
portfolio, here we deal with a portfolio with periodic cash installments $ c_t$. However, because
the latter are related to allocation decision variables by the constraint 
(\ref{constaint_rp}), the resulting 
quadratic reward (\ref{one_step_R_rp3}) has the same quadratic structure as the linear LQR reward (\ref{exp_R_LQR}). 
\end{comment}

\section{G-learner for retirement plan optimization}
\label{sect:G_learning_wealth}

To solve the optimization problem, we use a semi-analytical formulation of G-learning with Gaussian time-varying policies (GTVP).  
In what follows, we will refer to our specific algorithm implementing G-learning with our model specifications as the \emph{G-Learner} algorithm, to differentiate our model from more general models that could potentially be constructed using G-learning as a general RL method.

We start by specifying a functional form of the value function as a quadratic form of $ {\bf x}_t $:
\beq
 \label{F_parametrization_rp}
 F_t^{\pi}({\bf x}_t) =  {\bf x}_t^T {\bf F}_t^{(xx)}  {\bf x}_t  
 +   {\bf x}_t^{T} {\bf F}_t^{(x)} 
 + F_t^{(0)}, 
 \eeq 
where $ {\bf F}_t^{(xx)}, \, {\bf F}_t^{(x)}, \, F_t^{(0)} $ are parameters that can depend on time via their dependence on the target 
values $ \hat{P}_{t+1}$ and  
the expected returns $ \bar{\bf r}_t $. The dynamic equation takes the form:
\beq
\label{state_eq_LQR_rp}
{\bf x}_{t+1} = {\bf A}_t 
\left( {\bf x}_t + 
{\bf u}_t \right) 
 + \left( {\bf x}_t + 
{\bf u}_t \right) \circ % \varepsilon_{t}, 
 \tilde{\bf \varepsilon}_t, 
\; \; \; {\bf A}_t := 
\text{diag} \left(1 + \bar{\bf r}_t \right), \; \; 
 \tilde{\bf \varepsilon}_t := (0, {\bf \varepsilon}_t )
\eeq
%where
%\beq
%\label{A_B_LQR_rp}
%{\bf A}_t = 
%1 + r_f + \bar{\bf r}_t 
%\eeq
Note that the only features used here are the expected asset returns $ \bar{\bf r}_t $ for the current period $ t$. We assume that the expected asset returns are available as an output of a separate statistical model
using e.g. a factor model framework. The present formalism is agnostic to the choice of the expected return model.

Coefficients of the value function (\ref{F_parametrization_rp}) are computed backward in time starting from the last maturity $ t = T -1 $.
For $ t = T-1$, the quadratic reward 
(\ref{one_step_R_rp3})
can be optimized analytically by the following action:
\beq
\label{u_T_minus_1_rp}
{\bf u}_{T-1} = 
%\left( \hat{\Sigma}_{T-1} 
%+ \frac{\eta}{\lambda} I \right)^{-1}
\tilde{\bf \Sigma}_{T-1}^{-1}
\left( 
 \frac{1}{2 \lambda} {\bf R}_t^{(u)}
 + \frac{1}{2 \lambda} {\bf R}_t^{(ux)} {\bf x}_{T-1}
\right)
%\tilde{{\bf P}}_{T}  
%- \hat{ \bf \Sigma}_{T-1} {\bf x}_{T-1}  
%\right),
\eeq
where we defined  
$ \tilde{\bf \Sigma}_{T-1}$ 
%and $ {\bf \Phi} $   
%$  \tilde{{\bf P}}_{T} $ 
as follows
\beq
\label{short_not_rp}
\tilde{\bf \Sigma}_{T-1} := 
\hat{\bf \Sigma}_{T-1} + \frac{1}{\lambda}
{\bf \Omega}.
%, \; \; \; 
%{\bf \Phi} := \eta ( 1 + \bar{\bf r}_{T-1}) 
%{\bf 1}^T - \hat{\bf \Sigma}_{T-1}.
\eeq
Note that the optimal action is a linear function of the state.
Another interesting point to note is that the last term $ \sim {\bf \Omega} $ that describes convex transaction costs in Eq.(\ref{one_step_R_rp3}) produces regularization of matrix inversion in Eq.(\ref{u_T_minus_1_rp}). 

As for the last time step we have 
$ F_{T-1}^{\pi}({\bf x}_{T-1}) = \hat{R}_{T-1}$, coefficients  $ {\bf F}_{T-1}^{(xx)}, \, {\bf F}_{T-1}^{(x)}, \, F_{T-1}^{(0)} $ can be computed by plugging Eq.(\ref{u_T_minus_1_rp}) back in Eq.(\ref{one_step_R_rp3}), and comparing the result with 
Eq.(\ref{F_parametrization_rp}) with $ t = T-1$.
This provides terminal conditions for parameters in 
Eq.(\ref{F_parametrization_rp}):
\begin{eqnarray}
\label{term_conds_F_fun_rp}
{\bf F}_{T-1}^{(xx)} 
&=&  
{\bf R}_{T-1}^{(xx)} 
+ \frac{1}{2 \lambda}
\left[ {\bf R}_{T-1}^{(ux)} \right]^T 
\left[ \tilde{\bf \Sigma}_{T-1}^{-1} \right]^T
{\bf R}_{T-1}^{(ux)} 
+ \frac{1}{4 \lambda^2} 
\left[ {\bf R}_{T-1}^{(ux)} \right]^T 
\left[ \tilde{\bf \Sigma}_{T-1}^{-1} \right]^T
{\bf R}_{T-1}^{(uu)} \tilde{\bf \Sigma}_{T-1}^{-1}
{\bf R}_{T-1}^{(ux)} \nonumber \\
%- {\bf \Phi}^T \left[ \lambda \left( %\tilde{\bf \Sigma}_{T-1}^{-1} \right)^T %\hat{\bf \Sigma}_{T-1} \tilde{\bf %\Sigma}_{T-1}^{-1} + 
%{\bf \Omega} \right] {\bf \Phi} 
%- 2 \lambda \hat{\bf \Sigma}_{T-1} \tilde{\bf %\Sigma}_{T-1}^{-1} {\bf \Phi} 
%- 2 \lambda \left( \eta {\bf 1}(1 + \bar{\bf %r}_{T-1})^T + \frac{1}{2} \right) \hat{\bf %\Sigma}_{T-1} \nonumber \\
%&+& 2 \lambda \eta {\bf 1} (1 + \bar{\bf %r}_{T-1})^T + 2 \lambda \eta^2 {\bf 1} (1 + %\bar{\bf r}_{T-1})^T \tilde{\bf %\Sigma}_{T-1}^{-1} (1 + \bar{\bf r}_{T-1}) %{\bf 1}^T \nonumber \\
{\bf F}_{T-1}^{(x)} 
&=& 
{\bf R}_{T-1}^{(x)} 
%+ \frac{1}{2 \lambda} \tilde{\bf %\Sigma}_{T-1}^{-1} {\bf R}_{T-1}^{(u)}
+ \frac{1}{ \lambda} 
\left[{\bf R}_{T-1}^{(ux)} \right]^T 
\left[ \tilde{\bf \Sigma}_{T-1}^{-1} \right]^T
{\bf R}_{T-1}^{(u)}
+ \frac{1}{2 \lambda^2}  
\left[{\bf R}_{T-1}^{(ux)} \right]^T 
\left[ \tilde{\bf \Sigma}_{T-1}^{-1} \right]^T
{\bf R}_{T-1}^{(uu)} \tilde{\bf \Sigma}_{T-1}^{-1} {\bf R}_{T-1}^{(u)}
 \\
F_{T-1}^{(0)} 
&=& 
R_{T-1}^{(0)} + 
\frac{1}{2 \lambda}
\left[{\bf R}_{T-1}^{(u)} \right]^T 
\left[ \tilde{\bf \Sigma}_{T-1}^{-1} \right]^T
{\bf R}_{T-1}^{(u)} 
+ \frac{1}{4 \lambda^2}
\left[{\bf R}_{T-1}^{(u)} \right]^T 
\left[ \tilde{\bf \Sigma}_{T-1}^{-1} \right]^T
{\bf R}_{T-1}^{(uu)}  \tilde{\bf \Sigma}_{T-1}^{-1} 
{\bf R}_{T-1}^{(u)}
\nonumber.
%- \frac{1}{2 \lambda} {\bf 1}^T 
%\tilde{\bf \Sigma}_{T-1}^{-1} ( 1 + \bar{\bf %r}_{T-1})
%- \frac{1}{4 \lambda^2} 
%{\bf 1}^T \left[ \lambda \left( \tilde{\bf %\Sigma}_{T-1}^{-1} \right)^T \hat{\bf %\Sigma}_{T-1} \tilde{\bf \Sigma}_{T-1}^{-1} + 
%{\bf \Omega} \right] {\bf 1}
 \end{eqnarray}
For an arbitrary time step  $ t = T-2, \ldots, 0 $,  we use Eq.(\ref{state_eq_LQR_rp})
to compute the conditional 
expectation of the next-period F-function in the Bellman equation %(\ref{G_from_F_ch10_2}) 
as follows:
\begin{eqnarray}
\label{F_next_1_rp}
\mathbb{E}_{t, {\bf a}} \left[  F_{t+1}^{\pi}({\bf x}_{t+1})  \right] 
&=& 
\left( {\bf x}_t + {\bf u}_t \right)^T
\left( {\bf A}_t^T 
\bar{{\bf F}}_{t+1}^{(xx)}
{\bf A}_t + \tilde{\bf \Sigma}_r  \circ
\bar{{\bf F}}_{t+1}^{(xx)} \right) 
\left( {\bf x}_t + {\bf u}_t \right)
\nonumber \\
& +& \left( {\bf x}_t + {\bf u}_t \right)^T
{\bf A}_t^T
\bar{{\bf F}}_{t+1}^{(x)} + 
\bar{F}_{t+1}^{(0)}, \; \; \; 
\tilde{\bf \Sigma}_r  := 
 \left[ \begin{array}{cc}
 0   &  {\bf 0} \\
 {\bf 0} &  {\bf \Sigma}_r 
 \end{array} \right]
\end{eqnarray}
where $\bar{\bf F}_{t+1}^{(xx)} :=
\mathbb{E}_t \left[ {\bf F} _{t+1}^{(xx)} 
\right]$,
and similarly for $ \bar{{\bf F}}_{t+1}^{(x)}$
and $ \bar{F}_{t+1}^{(0)}  $. 
This is a quadratic function of $ {\bf x}_t $
and $ {\bf u}_t $, and has the same structure as the quadratic reward 
$ \hat{R} ({\bf x}_t, {\bf a}_t) $ in Eq.(\ref{one_step_R_rp3}).  
Plugging both expressions in the Bellman equation  
\[
G_t^{\pi} ( {\bf x}_t, {\bf u}_t) 
 = \hat{R}_{t}({\bf x}_{t}, {\bf u}_{t} ) 
 +  \gamma  \mathbb{E}_{t, {\bf u}} \left[  \left. F_{t+1}^{\pi} ( {\bf x}_{t+1}) \right|  {\bf x}_t, {\bf u}_t \right]
 \]
we see that the action-value function 
$ G_t^{\pi} ( {\bf x}_t, {\bf u}_t) $ should also be a quadratic 
function of $ {\bf x}_t $
and $ {\bf u}_t $:
\beq
\label{Q_fun_LQR_rp}
G_t^{\pi} ( {\bf x}_t, {\bf u}_t) =
{\bf x}_t^T {\bf Q}_t^{(xx)}  {\bf x}_t
+ {\bf u}_t^T {\bf Q}_t^{(ux)}  {\bf x}_t
+ {\bf u}_t^T {\bf Q}_t^{(uu)}  {\bf u}_t
+ {\bf x}_t^T {\bf Q}_t^{(x)} 
+ {\bf u}_t^T {\bf Q}_t^{(u)} + Q_{t}^{(0)},
%\left( {\bf x}_t + {\bf u}_t \right)^T
%{\bf Q}_t^{(xx)} 
%\left( {\bf x}_t + {\bf u}_t \right)
%+ \left( {\bf x}_t + {\bf u}_t \right)^T
%{\bf Q}_t^{(x)} + {\bf u}_t^T 
%{\bf Q}_t^{(uu)} {\bf u}_t + 
%{\bf u}_t^T {\bf Q}_t^{(u)} + 
%Q_{t}^{(0)}
\eeq
where
\bea
\label{Q_fun_coeffs_LQR_rp}
&& {\bf Q}_t^{(xx)}
 = 
% - \lambda \eta^2 {\bf 1} {\bf 1}^T
% + 2 \lambda \eta (1 + \bar{\bf r}_t) {\bf %1}^T
%- \lambda \hat{\bf \Sigma}_t 
 {\bf R}_t^{(xx)}
+ \gamma 
\left( {\bf A}_t^T 
\bar{{\bf F}}_{t+1}^{(xx)}
{\bf A}_t + \tilde{\bf \Sigma}_r \circ
\bar{{\bf F}}_{t+1}^{(xx)} \right) 
\nonumber \\
&&  {\bf Q}_t^{(ux)} = 
%2 \lambda \eta 
%(1 + \bar{\bf r}_t) {\bf 1}^T - 
%2 \lambda \hat{\bf \Sigma}_t 
{\bf R}_t^{(ux)}
+ 2 \gamma \left( {\bf A}_t^T 
\bar{{\bf F}}_{t+1}^{(xx)}
{\bf A}_t + \tilde{\bf \Sigma}_r \circ
\bar{{\bf F}}_{t+1}^{(xx)} \right) 
\nonumber \\
&& {\bf Q}_t^{(uu)}  = 
%- \lambda \hat{\bf \Sigma}_t 
 {\bf R}_t^{(uu)} 
+ \gamma 
\left( {\bf A}_t^T 
\bar{{\bf F}}_{t+1}^{(xx)}
{\bf A}_t + \tilde{\bf \Sigma}_r \circ
\bar{{\bf F}}_{t+1}^{(xx)} \right) 
- {\bf \Omega} \\
&& {\bf Q}_t^{(x)}
 = {\bf R}_t^{(x)} + 
 \gamma {\bf A}_t^T  \bar{{\bf F}}_{t+1}^{(x)}
\nonumber \\
&& {\bf Q}_t^{(u)} =  {\bf R}_t^{(u)}
+ \gamma {\bf A}_t^T  \bar{{\bf F}}_{t+1}^{(x)}\nonumber \\
&& Q_{t}^{(0)}  =  R_{t}^{(0)} + \gamma F_{t+1}^{(0)}. 
\nonumber 
\eea
%Note that $ {\bf Q}_t^{(xx)} $ and 
%$ {\bf Q}_t^{(x)} $ are, respectively, a quadratic % and linear functions of $ {\bf z}_t $.
%Note that the quadratic action-value function in Eq.(\ref{Q_fun_LQR_rp}) is similar to Eq.(\ref{Q_fun_LQR}) --- the only difference 
%is the specification of the parameters.

%Beyond different expressions for coefficients %of 
%the value function $ F_t(x_t) $ and %action-value function $ G_t(x_t, u_t)$ and a %different terminal condition, the rest of %calculations to perform one step of G-learning %is straightforward. 
After the action-valued function is computed as per Eqs.(\ref{Q_fun_coeffs_LQR_rp}), what remains is to compute the F-function
for the current 
step:
%can be found using 
%Eq.(\ref{F_opt_2}) repeated again here:
\beq
\label{F_opt_2_rp}
F_t^{\pi}({\bf x}_t ) =   \frac{1}{\beta} \log \int   \pi_0 
( {\bf u}_t | {\bf x}_t) e^{ \beta G_t^{\pi} ( {\bf x}_t, {\bf u}_t)  } d {\bf u}_t.
\eeq
A reference policy $ \pi_0 
( {\bf u}_t | {\bf x}_t) $ is Gaussian:
\beq
\label{Gaussian_pi_0_rp}
\pi_0 ( {\bf u}_t | {\bf x}_t) = 
\frac{1}{ \sqrt{\left( 2 \pi \right)^n \left| \Sigma_p \right| }}
e^{ - \frac{1}{2} \left( {\bf u}_t - 
\hat{\bf u}_t \right)^T \Sigma_p^{-1}
\left( {\bf u}_t - 
\hat{\bf u}_t \right) },
\eeq
where the mean value $\hat{\bf u}_t $ is a linear function of the state $ {\bf x}_t $:
\beq
\label{linear_mean_rp}
\hat{\bf u}_t = \bar{\bf u}_t + \bar{\bf v}_t 
{\bf x}_t.
\eeq
%Here $ \bar{\bf u}_t $ and $ \bar{\bf v}_t $ are %parameters that can be taken constants in the 
%prior distribution (\ref{Gaussian_pi_0_rp}). The %reason
%we keep the time label is that, as we will see %shortly, the optimal policy obtained from %G-learning
%with linear dynamics (\ref{state_eq_LQR})
%is also a Gaussian that can be written in the %same form as (\ref{Gaussian_pi_0_rp}) but with %updated parameters $ \bar{\bf u}_t $ and $ %\bar{\bf v}_t $ that will become time-dependent %due to their dependence on the targets $ %\hat{P}_t $ and expected asset returns $ \bar{\bf %r}_t $.

%Again as in Sect.~\ref{sect:Zero_friction},
Integration over $ {\bf u}_t $ in Eq.(\ref{F_opt_2_rp}) is performed analytically using the well known $ n$-dimensional Gaussian integration formula
\beq
\label{N_dim_Gauss}
\int e^{ - \frac{1}{2} {\bf u}^T {\bf A} {\bf u} + 
{\bf u}^T {\bf B}} d^n {\bf u} =
\sqrt{ \frac{ (2 \pi)^n}{ \left| {\bf A} \right| }}
e^{ \frac{1}{2} {\bf B}^T {\bf A}^{-1} {\bf B}},
\eeq
where $  \left| {\bf A} \right| $ denotes the determinant of matrix $ {\bf A} $.

Note that, unlike in the Merton approach \citep{Merton_1971} or in traditional Markowitz portfolio optimization \citep{Markowitz}, here we work with unconstrained variables that do not have to sum up to one, and therefore an unconstrained multivariate Gaussian integration readily applies here.
Remarkably, this implies that once the decision variables are chosen appropriately, portfolio optimization for wealth management tasks may in a sense be an easier problem than portfolio optimization that does not involve intermediate cashflows, and is often formulated using self-financing conditions.

%If no constraints are imposed on $ {\bf u}_t $, 
% with a Gaussian reference policy 
% $ \pi_0 $ 
%can be easily performed analytically as long as %$G_t^{\pi} ( {\bf x}_t, {\bf u}_t) $
%is quadratic in $ {\bf u}_t $. This is done using %the 
%$ n $-dimensional Gaussian integration formula
%\beq
%\label{N_dim_Gauss_rp}
%\int e^{ - \frac{1}{2} {\bf x}^T {\bf A} {\bf x} %+ 
%{\bf x}^T {\bf B}} d^n {\bf x} =
%\sqrt{ \frac{ (2 \pi)^n}{ \left| {\bf A} \right| %}}
%e^{ \frac{1}{2} {\bf B}^T {\bf A}^{-1} {\bf B}}
%\eeq
%where $  \left| {\bf A} \right| $ stands for the %determinant of matrix $ {\bf A} $. Using this %relation to calculate the integral in %Eq.(\ref{F_opt_2_rp}) and 

Performing the Gaussian integration and comparing the resulting expression with Eq.(\ref{F_parametrization_rp}), we 
obtain for its coefficients: 
%find that the resulting $ F$-function has the %same structure as in , where the coefficients are %now {\it computed} in 
%terms of coefficients of the $ Q$-function (see 
%exercise 10.3):
\bea
\label{F_fun_coeffs_LQR_rp}
F_t^{\pi}({\bf x}_t) 
&=& 
 {\bf x}_t^T {\bf F}_t^{(xx)}  {\bf x}_t  
 +   {\bf x}_t^{T} {\bf F}_t^{(x)} 
 + F_t^{(0)}  \nonumber \\
{\bf F}_t^{(xx)}
&=& {\bf Q}_t^{(xx)}
+ \frac{1}{2 \beta} \left(
{\bf U}_t^T \bar{\bf \Sigma}_p^{-1} {\bf U}_t
- \bar{\bf v}_t^T {\bf \Sigma}_p^{-1} \bar{\bf v}_t
\right) \nonumber \\
{\bf F}_t^{(x)}
&=& {\bf Q}_t^{(x)}
+ \frac{1}{\beta} \left(
{\bf U}_t^T \bar{\bf \Sigma}_p^{-1} {\bf W}_t
- \bar{\bf v}_t^T {\bf \Sigma}_p^{-1} \bar{\bf u}_t
\right)  \\
{\bf F}_t^{(0)}
&=& {\bf Q}_t^{(0)}
+ \frac{1}{2 \beta} \left(
{\bf W}_t^T \bar{\bf \Sigma}_p^{-1} {\bf W}_t
- \bar{\bf u}_t^T {\bf \Sigma}_p^{-1} \bar{\bf u}_t
\right)  - \frac{1}{2 \beta} \left( \log 
\left| {\bf \Sigma}_p \right| + \log \left|
\bar{\bf \Sigma}_p \right| \right), 
\nonumber  
%{\bf F}_t^{(xx)}
%\hskip-0.5cm &&= {\bf Q}_t^{(xx)}
%+ \frac{1}{2 \beta} \left(
%{\bf U}_t^T {\bf A}_t^{-1} {\bf U}_t
%+ \bar{\bf v}_t^T {\bf \Sigma}_p^{-1} \bar{\bf %v}_t
%\right) \nonumber \\
%{\bf F}_t^{(x)}
%\hskip-0.5cm &&= {\bf Q}_t^{(x)}
%+ \frac{1}{\beta} \left(
%{\bf U}_t^T {\bf A}_t^{-1} {\bf W}_t
%+ \bar{\bf v}_t^T {\bf \Sigma}_p^{-1} \bar{\bf %u}_t
%\right)  \\
%{\bf F}_t^{(0)}
%\hskip-0.5cm &&= {\bf Q}_t^{(0)}
%+ \frac{1}{2 \beta} \left(
%{\bf W}_t^T {\bf A}_t^{-1} {\bf W}_t
%+ \bar{\bf u}_t^T {\bf \Sigma}_p^{-1} \bar{\bf %u}_t
%\right)  - \frac{1}{2 \beta} \text{Tr} \log 
%\frac{ {\bf A}_t}{ \Sigma_p^{-1}}
%\nonumber 
\eea
where we use the auxiliary parameters
\bea
\label{aux_coeffs_LQR_rp}
{\bf U}_t 
&=&
\beta {\bf Q}_t^{(ux)} + \Sigma_p^{-1}
\bar{\bf v}_t \nonumber \\
{\bf W}_t 
&=&
\beta {\bf Q}_t^{(u)} + \Sigma_p^{-1}
\bar{\bf u}_t \\
\bar{\bf \Sigma}_p &=& {\bf \Sigma}_p^{-1} - 2 \beta
{\bf Q}_t^{(uu)} \nonumber. 
\eea
The optimal policy for the given step is given by
%can be found using Eq.(\ref{pi_opt_F_ch10_2}) repeated again here:
\beq
\label{pi_opt_F_ch10_2_rp}
 \pi ( {\bf u}_t | {\bf x}_t) = \pi_0 ( {\bf u}_t | {\bf x}_t) e^{ \beta \left(G_t^{\pi} ( {\bf x}_t, {\bf u}_t) - F_t^{\pi}({\bf x}_t ) 
  \right) }. 
 \eeq 
Using here the quadratic action-value function 
(\ref{Q_fun_LQR_rp}) produces 
a new Gaussian policy $  \pi ( {\bf u}_t | {\bf x}_t) $:
\beq
\label{pi_opt_LQR_rp}
\pi ( {\bf u}_t | {\bf x}_t) = 
\frac{1}{ \sqrt{\left( 2 \pi \right)^n \left| 
\tilde{\Sigma}_p \right| }}
e^{ - \frac{1}{2} \left( {\bf u}_t - 
\tilde{\bf u}_t - \tilde{\bf v}_t {\bf x}_t \right)^T \tilde{\Sigma}_p^{-1}
\left( {\bf u}_t - 
\hat{\bf u}_t - \tilde{\bf v}_t {\bf x}_t \right) }
\eeq
where
\bea
\label{Bayesian_update_LQR_rp}
\tilde{\Sigma}_p^{-1}
& =&
{\bf \Sigma}_p^{-1} - 2 \beta 
{\bf Q}_t^{(uu)} 
%\Sigma_p^{-1} + \beta 
%\left( {\bf Q}_t^{(xx)} + {\bf Q}_t^{(uu)} \right)
\nonumber \\
\tilde{\bf u}_t & =&
\tilde{\Sigma}_p \left( {\bf \Sigma}_p^{-1} 
\bar{\bf u}_t +  \beta {\bf Q}_t^{(u)} \right) \\
 %\tilde{\Sigma}_p \left( \Sigma_p^{-1} \bar{\bf %u}_t
 %- \beta {\bf Q}_t^{(x)} \right) \\
\tilde{\bf v}_t & =&
\tilde{\Sigma}_p \left( {\bf \Sigma}_p^{-1} 
\bar{\bf v}_t +  \beta {\bf Q}_t^{(ux)} \right)
\nonumber
 %\tilde{\Sigma}_p \left( \Sigma_p^{-1} \bar{\bf %v}_t
 %- \beta {\bf Q}_t^{(xx)} \right) \nonumber 
\eea
Therefore, policy optimization 
for G-learning with quadratic rewards and Gaussian reference policy amounts to the Bayesian 
update of the prior distribution 
(\ref{Gaussian_pi_0_rp}) with parameters 
updates 
$ \bar{\bf u}_t, \, \bar{\bf v}_t, \, \Sigma_p $ to
the new values $ \tilde{\bf u}_t, \, \tilde{\bf v}_t, \, \tilde{\Sigma}_p $ defined in Eqs.(\ref{Bayesian_update_LQR_rp}). These quantities 
depend on time via their dependence on 
the targets $ \hat{P}_t $ and expected asset returns $ \bar{\bf r}_t $.

%As in  Sect.~\ref{sect:Zero_friction}, 
For a given time step $ t $, the G-learning 
algorithm keeps iterating between the 
policy optimization step that updates policy parameters according to Eq.(\ref{Bayesian_update_LQR_rp}) for fixed coefficients of the $ F$- and $ G$-functions, 
and the policy evaluation step that involves Eqs.(\ref{Q_fun_LQR_rp}, \ref{Q_fun_coeffs_LQR_rp},
\ref{F_fun_coeffs_LQR_rp}) and solves for parameters 
of the $ F$- and $ G$-functions given policy parameters. Note that convergence of iterations for 
$ \tilde{\bf u}_t, \, \tilde{\bf v}_t$ is guaranteed as $ \left|\tilde{\Sigma}_p {\bf \Sigma}_p^{-1} \right| < 1 $.
At convergence of iteration for time step $ t$, 
Eqs.(\ref{Q_fun_LQR_rp}, \ref{Q_fun_coeffs_LQR_rp},
\ref{F_fun_coeffs_LQR_rp}) and (\ref{pi_opt_LQR_rp}) together solve one step of G-learning.
The calculation then proceeds by moving to the previous step $ t \rightarrow t - 1 $, and
repeating the calculation, all the way back to the present time.

The additional step needed from G-learning for the present problem is to find the optimal cash contribution for each time step by using the budget constraint (\ref{constaint_rp}). As G-learning produces Gaussian random actions $ {\bf u}_t$, Eq.(\ref{constaint_rp}) implies that the time-$t$ optimal contribution $ c_t$ is Gaussian distributed with mean  $ \bar{c}_t = \mathbbm{1}^T \left(\bar{\bf u}_t +  \bar{\bf v}_t {\bf x}_t \right)$. 
The expected optimal contribution $  \bar{c}_t $ thus has a part 
$ \sim \bar{\bf u}_t $ that is independent
of the portfolio value, and a part $ \sim \bar{\bf v}_t $ that depends on the current portfolio. 
This is similar e.g. to a linear specification of the defined contribution with a deterministic policy in \cite{Lin_Zeng_2019}.
%At each of ten periods, the mean optimal contribution %estimated using G-learning is shown in Figure %\ref{fig:G-cash}.
%and variance $ N \bar{\bf \Sigma}_p $. 

It should be noted that in practice, we may want to impose constraints on cash installments $ c_t$. For example, we could impose band constraints $ 0 \leq c_t \leq c_{max} $ with some upper bound $ c_{max}$.
Such constraints can be easily added to the framework. To this end, we need to replace 
the exactly solvable unconstrained least squares problem with a constrained least squares problem. This can be done without a substantial increase of computational time using efficient off-the-shell convex optimization software. Note that 
enforcing constraints on the resulting cash-flows in our approach amounts to optimization with one constraint, instead of two constraints as in the Merton approach.

\section{GIRL: G-learning IRL}
\label{sect_GIRL}

So far in this paper, we considered the setting of (direct) reinforcement learning, when the agent (investor) learns while observing the rewards, and optimizes the policy so that the expected cumulative reward (regularized by the KL information cost) is maximized. This setting is suitable when the investor explicitly defines his or her reward function.

In many cases of practical interest, 
an individual investor may not be able to explain his or her utility function used for trading decision-making, which can instead be rule-driven (or driven by other model not formulated in RL terms). Alternatively, when an agent (investor) is a subject of behavioral inference to a different agent (a researcher or robo-advisor), the latter has access to observed trajectories (states and actions) of the agent, but not to rewards received by the agent.
Such cases where rewards are not available belong in the realms of Inverse Reinforcement Learning (IRL) whose objective is to recover \emph{both} the reward function of the agent and the optimal policy, see e.g. \citep{DHB} for a review.

In this section, we consider the IRL problem with G-learning, and present an algorithm we call GIRL (G-learning IRL) whose objective is to 
make inference of the reward function of an individual agent such as a retirement plan contributor or an individual brokerage account holder.
That is, we assume that we are given a history of  dollar-nominated asset positions in an investment portfolio, jointly with an agent's decisions that include both injections or withdrawals of cash from the portfolio and asset allocation decisions. Additionally, we are given historical values of asset prices and expected asset returns for all assets in the investor universe. As previously in the paper, we can consider a portfolio of stocks and a single bond, but the same formalism can be applied to other types of assets.

Assume that we have historical data that includes a set of $ D $ trajectories 
$ \zeta_i $ where $ i = 1, \ldots D $ of state-action pairs $ ( {\bf x}_t, {\bf u}_t ) $ where trajectory $ i $ starts at some time $ t_{0i} $ and runs until time
 $ T_i $. 
Consider a single trajectory $ \zeta $ from this collection, and set for this trajectory the start time $ t = 0 $ and the end time $ T $. As individual trajectories are considered independent, they 
will enter additively in the final log-likelihood of the problem. We assume that dynamics are Markovian in the pair $ ( {\bf x}_t,  {\bf u}_t ) $, with a generative model 
$ p_{\theta} ({\bf x}_{t+1}, {\bf u}_t  |  {\bf x}_t) =  \pi_{\theta} ( {\bf u}_t | {\bf x}_t)  p_{\theta} \left( {\bf x}_{t+1} | {\bf x}_t, {\bf u}_t  \right) $ where 
$ \Theta $ stands for a vector of model parameters, and $ \pi_{\theta} $ is the action policy given by Eq.(\ref{pi_opt_F_ch10_2_rp}).

The probability of observing trajectory $ \zeta $ is  given by the following expression
\beq
\label{P_c}
P \left( {\bf x}, {\bf u} | \Theta \right) =  p_{0}( {\bf x}_0) \prod_{t =0}^{T-1} \pi_{\theta}  
( {\bf u}_{t} | {\bf x}_{t}  ) p_{\theta} \left( {\bf x}_{t+1} | {\bf x}_t,  {\bf u}_t  \right). 
 \eeq
 Here $ p_{0}( {\bf x}_0) $ is a marginal probability of $ {\bf x}_t $ at the start of the $ i$-th demonstration.
Assuming that the initial values $ {\bf x}_0$ are fixed, 
this gives the following log-likelihood for data 
$ \left\{ {\bf x}_t, {\bf a}_t \right\}_{t=0}^{T} $ observed for trajectory $ \zeta $: 
 \beq
 \label{full_LL}
 LL ( {\bf \theta})  :=  
\log P \left( {\bf x}, {\bf u} | \Theta \right) = 
%\log  p_{\theta}( {\bf x}_0) + 
\sum_{t \in \zeta} \left( \log  \pi_{\theta} ( {\bf u}_t | {\bf x}_t) + 
\log p_{\theta} \left( {\bf x}_{t+1} | {\bf x}_t,  {\bf u}_t  \right) \right).
\eeq
Transition probabilities $ p_{\theta} \left( {\bf x}_{t+1} | {\bf x}_t,  {\bf u}_t  \right) $ entering this expression can be obtained from the state equation 
\beq
\label{state_eq_LQR_ch11}
{\bf x}_{t+1} = {\bf A}_t 
\left( {\bf x}_t + 
{\bf u}_t \right) 
 + \left( {\bf x}_t + 
{\bf u}_t \right) \circ % \varepsilon_{t}, 
 \tilde{\bf \varepsilon}_t, 
\; \; \; {\bf A}_t := 
\text{diag} \left(1 + \bar{\bf r}_t \right), \; \; 
 \tilde{\bf \varepsilon}_t := (0, {\bf \varepsilon}_t ),
\eeq
where $ \varepsilon_t $ is a Gaussian noise with covariance $ {\bf \Sigma}_r $ (see 
Eq.(\ref{state_eq_LQR_rp})). Writing $ {\bf x}_t = (x_t^{(0)}, {\bf x}_t^{(r)} )$ where $ x_t^{(0)} $ is the value of a bond position and $ {\bf x}_t^{(r)} $ are the values of positions in risky assets, and similarly for 
$ {\bf u}_t $ and $ {\bf A}_t$,
this produces transition probabilities
\beq
\label{one_step_p_x}
p_{\theta} \left( {\bf x}_{t+1} | {\bf x}_t, {\bf u}_t \right) = \frac{e^{ - \frac{1}{2} {\bf \Delta}_{t }^{T} {\bf \Sigma}_r^{-1} 
{\bf \Delta}_{t }}}{\sqrt{ \left( 2 \pi \right)^{N} \left| {\bf \Sigma}_r \right|}}  \delta \left( x_{t+1}^{(0)} - (1+r_f) 
x_{t}^{(0)} \right), \; \; {\bf \Delta}_{t }
 := \frac{ {\bf x}_{t+1}^{(r)}}{ {\bf x}_t^{(r)} + {\bf u}_t^{(r)}} - 
\vec{\bf A}_t^{(r)},  
\eeq
where the factor $ \delta \left( x_{t+1}^{(0)} - (1+r_f) 
x_{t}^{(0)} \right)$ captures the deterministic dynamics of the bond part of the portfolio. As this term does not depend on model parameters, we can drop it from the log-transition probability, along with a constant term
$ \sim \log(2 \pi)$. This produces
\beq
\label{log_trans_prob_ch11}
\log p_{\theta} \left( {\bf x}_{t+1} | {\bf x}_t, {\bf u}_t \right) = - \frac{1}{2} \log \left| {\bf \Sigma}_r \right| - \frac{1}{2} {\bf \Delta}_{t }^{T} {\bf \Sigma}_r^{-1} 
{\bf \Delta}_{t }.
\eeq
Substituting Eqs.(\ref{pi_opt_F_ch10_2_rp}), (\ref{Q_fun_LQR_rp}), (\ref{log_trans_prob_ch11}) into the trajectory log-likelihood (\ref{full_LL}), we put it in the following form:
\beq
 \label{full_LL_2}
 LL ( {\bf \theta}) = 
%\log  p_{\theta}( {\bf x}_0) + 
\sum_{t \in \zeta}  \left(  \beta \left( G_t^{\pi} ( {\bf x}_t, {\bf u}_t) - F_t^{\pi}({\bf x}_t ) 
\right)
 - \frac{1}{2} \log \left| {\bf \Sigma}_r \right| - \frac{1}{2} {\bf \Delta}_{t }^{T} {\bf \Sigma}_r^{-1} 
{\bf \Delta}_{t } 
\right),
\eeq
where $ G_t^{\pi} ( {\bf x}_t, {\bf u}_t) $ and 
$ F_t^{\pi}({\bf x}_t ) $ are defined by Eqs.(\ref{Q_fun_LQR_rp}) and (\ref{F_parametrization_rp}). 
The log-likelihood (\ref{full_LL_2}) is a function of model parameter vector $ {\bf \theta} = \left( \lambda, \eta, \rho, {\bf \Omega},
{\bf \Sigma}_r, {\bf \Sigma}_p,
\bar{\bf u}_t,\bar{\bf v}_t \right) $
(recall that $ \beta $ is a regularization hyper-parameter which should not be optimized in-sample). 
We can simplify the problem by setting 
$ \bar{\bf v}_t = 0$ and $ \bar{\bf u}_t = \bar{\bf u} $ (i.e. take a constant mean in the prior). 
In this case, the vector of model parameter
to learn with IRL inference is $ {\bf \theta} = \left( \lambda, \eta, \rho, {\bf \Omega}, 
{\bf \Sigma}_r, {\bf \Sigma}_p, 
\bar{\bf u} \right) $.
A ``proper'' IRL setting would correspond
to only learning parameters of the reward function   $ \left( \lambda, \eta, \rho, {\bf \Omega} \right) $ while keeping parameters $ \left({\bf \Sigma}_r, {\bf \Sigma}_p, 
\bar{\bf u} \right)$ fixed (i.e. estimated outside of the IRL model).
Optimization can be performed using available off-the-shelf software. In our implementation, we use the Adam optimization method within PyTorch to optimize the negative log-likelihood function.

\section{Numerical examples} \label{sect:num}
% emphasize that G-learning/GIRL doesn't assume a DGP
To illustrate the G-learner and GIRL algorithms for goal based wealth management, we use a simple
simulated environment that mimics the working of equity return models (sometimes referred to as ``alpha-models'') which are expected in practice to be weak predictors of realised returns. The advantage of such a simulated environment is that it allows us to define the ``ground truth'' and thus demonstrate the performance of both algorithms. We remind the reader that while we use simulated data to show the performance of our algorithms, the latter are \emph{model free} as they are independent of a model of stock-price dynamics.

The investment horizon is set to 7.5 years and the portfolio rebalancing and consumption occur quarterly (over 30 periods). In this simplified setting, the portfolio is assumed to be initially equally weighted, with \$1000 allocated equally between $N-1=99$ stocks and a risk free bond. We assume a fixed risk free annual rate, $r_f=0.02$, stock transactions costs are 1.5\% of the stock price and risk-free bond transactions costs are 5\%.
%The target portfolio is 
%initially set to the initial value of the %initial client portfolio times 1.1, and then %continuously compounds at a constant rate of %15\%. 
The benchmark portfolio is initially set equal to the initial value of the portfolio, and is continuously compounded at a constant rate of 50\%. 
%For the specific choice of model parameters %used in this example, the model optimally %chooses to initially invest a large amount %into the plan and withdraw funds later in %the plan's lifetime.

We model the quarterly realized risky asset returns, $r_{t,i}$, of the $i^{th}$ asset as being correlated to expected risky asset returns, $\bar{r}_{t,i} $:
\beq
r_{t,i} = \bar{r}_{t,i} + \beta'_i (r_M - \mu_Mdt)  
                         + \sigma_i \sqrt{1 - (\beta'_i)^2} dW_{t,i}, ~i \in \{1, \dots, N-1\},
\eeq
where $\mu_M=0.05$ is the market drift, $r_M$ are the market returns simulated under a GBM model with volatility $\sigma_M=0.25$, and $\beta'_i$ is the beta of the $i^{th}$ asset. $\sigma_i\equiv\sigma=0.05$ is the idiosyncratic volatility and $dW_t$ is a driving Brownian motion which is correlated with the market noise and $dt=0.25$.  $\bar{r}_t$ is assumed to be given by CAPM: % with a dynamic market risk premium:
\beq
\bar{r}_t= \alpha + \beta' ((1-c)\mu_Mdt  + cr_M), ~c\in[0,1]
\eeq
where we choose the oracle coefficient $c=0.2$.

\begin{figure}[H]
    \centering
    \includegraphics[width=0.5
    \textwidth]{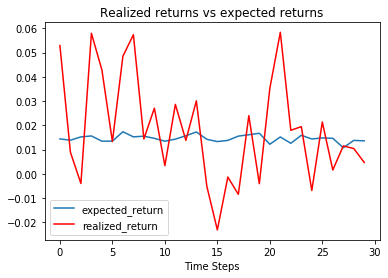}
    \caption{\textit{The sample mean realized returns are plotted against the sample mean expected returns and observed to be weakly correlated.}}
    \label{fig:exp_returns}
\end{figure}

We assume that $\alpha$ and $\beta'$ are uniform random variables across all risky assets, with $\alpha\sim \mathcal{U}([-0.05,0.15]), ~ \beta'\sim \mathcal{U}([0.05,0.85])$. The risky assets are assumed to initially be dollar values given by uniform random variables $\mathcal{U}([20,120])$. In our experiments, we generate the risky asset returns over $M=1000$ paths using sampling noise under i.i.d. Gaussian vector distributions. Figure \ref{fig:exp_returns} compares the sample mean of the simulated realized returns with the sample mean of the expected returns, which are observed to be weakly correlated.

To demonstrate a G-learning agent for wealth management, we arbitrarily choose the set of parameters in Table \ref{tab:params}. Note that the G-learner parameter, $\beta$, is not optimized by GIRL, but is simply set as $\beta=1000$ to ensure numerical stability in the G-learner. In practice $\beta>0$ can be chosen arbitrarily in GIRL without affecting its ability to learn the rewards from state-action trajectories, although the learning behavior is changed (see Section \ref{sect:ent-bellman}). 

\begin{figure}[H]
    \centering
    \includegraphics[width=0.5
    \textwidth]{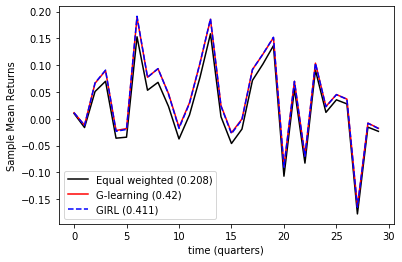}
    \caption{\textit{The sample mean portfolio returns are shown over a 30 quarterly period horizon (7.5 years). The black line shows the sample mean returns for an equally weighted portfolio without rebalancing. The red line shows a G-learning agent, for the parameter 
    values given in Table \ref{tab:params}. GIRL imitates the G-learning agent and generates returns shown by the blue dashed line. Sharpe Ratios are shown in parentheses. }}
    \label{fig:sharpe}
\end{figure}

The G-learner takes as input the expected risky asset returns $\bar{r}_t$ together with the covariance of the risk asset return, $\Sigma_r$. The discount factor for the future value of rewards, $\gamma=0.95$. As shown in Figure \ref{fig:sharpe}, even using these arbitrary parameters results in superior Sharpe ratios when compared with an equally weighted portfolio that is never rebalanced over the investment horizon. The G-learner uses the alpha-model to consistently produce superior returns in a multi-period setting using a locally-quadratic reward function. The G-learner trains in a few seconds on a portfolio of 100 assets on standard hardware.

\begin{figure}[H]
  \centering
\begin{tabular}{cc}
  
    \includegraphics[width=0.45
    \textwidth]{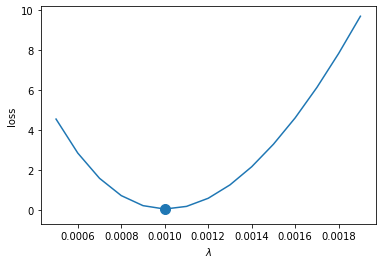} &
    \includegraphics[width=0.45
    \textwidth]{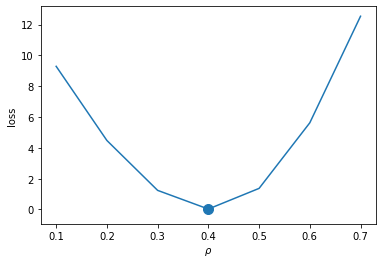}\\
    (a) $\lambda$ & (b) $\rho$ \\
    \includegraphics[width=0.45
    \textwidth]{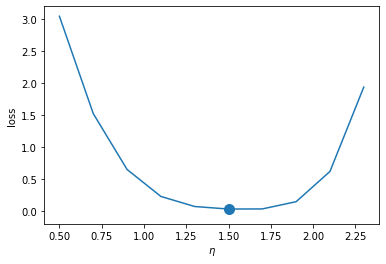} &
    \includegraphics[width=0.45
    \textwidth]{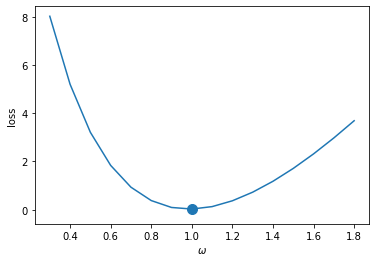}\\
     (c) $\eta$ & (d) $\omega$\\
    
\end{tabular}    
    \caption{\textit{The loss surface about each of the G-learner's parameters which are found by GIRL. The solid circle denotes the exact parameter value. The loss is convex w.r.t. to each parameter. }}
    \label{fig:loss}
\end{figure}

GIRL imitates the G-learner by minimizing a loss function over the state-action trajectories generated by the G-learner. The GIRL learned parameters in Table $\ref{tab:params}$ are observed to be close to the G-learner parameters up to sampling error and numerical accuracy. GIRL is implemented using the ADAM method for stochastic gradient descent with a learning rate $\ell=0.1$ and a stopping tolerance on the parameter vector, $\tau=1\times 10^{-8}$. Consequently GIRL is observed to imitate the G-learner --- the sample averaged portfolio returns closely track each other in Figure \ref{fig:sharpe}. The error in the learned G-learner parameters results in a marginal decrease in the Sharpe ratio, as reported in the parentheses of the legend in Figure \ref{fig:sharpe}. In Figure \ref{fig:loss}, we show the local behaviour of the loss surface for our problem, illustrating its convex shape and parameters found by GIRL. GIRL requires approximately 200 iterations to converge.

\begin{table}[H]
    \centering
    \begin{tabular}{c|cc}
    \hline 
    Parameter & G-learner & GIRL\\
    \hline
       $\rho$  &  0.4 &  0.406\\
        $\lambda$ & 0.001& 0.000987\\
        $\eta$ & 1.01 & 1.0912\\
        $\omega$ & 0.15& 0.149 \\
        \hline
    \end{tabular}
    \caption{\textit{The G-learning agent parameters used for portfolio allocation together with the values estimated by GIRL.}}
    \label{tab:params}
\end{table}

\begin{figure}
  \centering
  \ifdefined\final
  \includegraphics[width=0.7
    \textwidth]{figures/eps/G_learning_wealth_unconstrained.eps}\\
   
  \else
  {
    \includegraphics[width=0.5
    \textwidth]{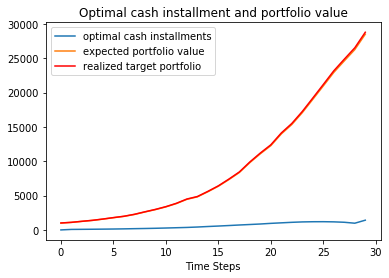}\\
    }
    \fi
\caption{\textit{
An illustration of the G-learner for a retirement plan optimization using a portfolio with 100 assets. 
The values of optimal cash installments are shown in Table~\ref{tab:my_label}.
%If no additional constraints on cash installments are imposed and for certain market scenarios and model parameters, unconstrained optimization may produce solutions where it is optimal to deposit a large amount at the beginning, and then withdraw funds later in the plan's lifetime
%At each of ten periods, the optimal contribution %estimated using G-learning is a Gaussian distribution %with mean $ \bar{c}_t = \mathbbm{1}^T \left(\bar{\bf u}_t %+  \bar{\bf v}_t {\bf x}_t \right)$. The optimal mean %investment is observed here to monotonically decrease %in time.
}}
      \label{fig:G-cash}
\end{figure}

An illustration of 
an optimal solution trajectory obtained without enforcing any constraints is shown in Figure \ref{fig:G-cash} which presents simulation results for the portfolio using
the G-learner. The values of optimal cash installments are shown in Table~\ref{tab:my_label}. %We observe the %If no additional constraints on cash installments are imposed and for certain market scenarios and model parameters, unconstrained optimization may produce solutions where it is optimal to deposit a large amount at the beginning, and then withdraw funds later in the plan's lifetime

\begin{table}[H]
    \centering
    \begin{tabular}{c|c}
    \hline
    Period & Expected Cash Installments (\$) \\
    \hline
     1&0.0\\
2&73.384\\
3&85.7\\
4&97.36\\
5&113.083\\
6&129.889\\
7&153.362\\
8&181.832\\
9&207.472\\
10&237.292\\
11&275.926\\
12&318.154\\
13&360.212\\
14&420.546\\
15&495.813\\
16&563.691\\
17&638.042\\
18&716.391\\
19&787.57\\
20&861.794\\
21&954.392\\
22&1030.161\\
23&1106.024\\
24&1164.276\\
25&1190.959\\
26&1196.982\\
27&1173.541\\
28&1112.945\\
29&976.385\\
30&1416.265\\
\hline
    \end{tabular}
    \caption{Optimal cash installment for the portfolio process shown in Figure \ref{fig:G-cash}.}
    \label{tab:my_label}
\end{table}

%\section{GIRL: G-Inverse Reinforcement Learning}
%\label{sect:GIRL}
%All results are obtained from our %implementation of G-learning and GIRL in %PyTorch (version 1.4.0). 

\section{Summary} \label{sect:concl}

To summarize, in this paper we presented a reinforcement learning (RL) based approach to problems of wealth management such as retirement plans. We used a generative 
framework for RL known as G-learning, and developed its practical implementation for both problems of optimization the policy given rewards (direct RL), and the inverse problem of finding the reward function of an agent from its observed behavior (inverse RL, or IRL). This resulted in two related practical algorithms that we called G-Learner and GIRL.   

Our approach is applicable provided we use absolute (dollar-nominated) asset position changes as action variables, and choose a reward function which is quadratic in these actions.
As shown in 
Sect.~\ref{sect:G_learning_wealth}, 
G-learning with a quadratic reward and Gaussian reference policy gives rise to an entropy-regulated LQR as a novel tool for wealth management tasks. This approach results in 
a Gaussian optimal policy whose mean is a linear function of the state
$ {\bf x}_t $. 

The method we presented enables extensions to other formulations including constrained versions or other specifications of the reward function.
One possibility is to use the definition 
 in Eq. (\ref{one_step_R_rp}) with the constraint in Eq. (\ref{constaint_rp})), which provides an example of a non-quadratic concave reward.
Such cases should be implemented using flexible function approximations for the action-value function such as neural networks. 

By focusing on a semi-analytically tractable G-learning based approach to goal-based wealth management, we presented two practical algorithms that we called G-Learner and 
GIRL. As we showed using simulations where the ``ground truth'' is known, G-Learner is able to improve over the benchmark equally-weighted portfolio strategy, while GIRL is able to successfully recover parameters of an agent which is modeled as a G-Learner. 

Given that behavioral data generated in our approach are very noisy (as it also happens in real financial markets), a success of such an endeavour could not be guaranteed beforehand, at neither stage. Indeed, the very ability of G-Learner to perform better than the benchmark equally-weighted portfolio is hinged, as could be expected, on the ability of the equity return model (the ``alpha-model'') to exhibit some (rather weak) predictive power. Unlike a passive manager of the equally-weighted portfolio, the G-Learner is able to harvest the predictive power of the alpha model, providing a consistent boost in terms of resulting Sharpe ratio.  Our numerical experiments demonstrate that the G-learner uses the alpha-model to consistently produce superior returns in a multi-period setting using a locally-quadratic reward function.

Furthermore, strong statistical noise in the data could also render the inverse problem of inference of the reward function of a G-Learner agent very difficult. As we demonstrated with experiments, however, GIRL manages 
to imitate the G-Learner, i.e. it infers the correct reward parameters, and thus imitates a G-Learner.

The two algorithms, G-Learner and GIRL, can be used either separately or in a combination. In particular, their combination could be used in robo-advising by modeling the actual human agents as G-learners, and then use GIRL to infer the latent objectives (rewards) of these G-learners. GIRL would then be able to imitate the best human investors, and thus could be offered as a robo-advising service to clients that would allow them to perform on par with best performers among all investors.

\bibliographystyle{chicago}
\bibliography{master}

\end{document}